\documentclass{article}
\usepackage{amsmath}
\usepackage{graphicx}
\usepackage{latexsym}
\usepackage{amssymb}
\usepackage{multirow}
\usepackage{array}
\usepackage{color}
\usepackage{epsfig} 
\newtheorem{theorem}{Theorem}

\newtheorem{lemma}[theorem]{Lemma}

\newtheorem{definition}[theorem]{Definition}
\newtheorem{proposition}[theorem]{Proposition}
\newtheorem{example}[theorem]{Example}
\newtheorem{remark}[theorem]{Remark}

 \newcommand{\R}{\mathbb{R}} 
   
 \newcommand{\Rb}{{\overline{\mathbb{R}}}}

\newcommand{\lfteqn}{\vspace{-0.08in} \begin{eqnarray} \begin{array}{lllllll}}
\newcommand{\ndeqn}{\vspace{-0.08in} \end{array} \nonumber \end{eqnarray}}
\newcommand{\Lfteqn}{\vspace{-0.08in} \begin{eqnarray} \begin{array}{lllllll}}
\newcommand{\Ndeqn}{\vspace{-0.08in} \end{array}  \end{eqnarray}}

\title{\bf Analysis of P-time Event Graphs in (Max,+) and (Min,+) Algebras}

\author{Pavel \v{S}pa\v{c}ek, Jan Komenda, and S\'ebastien Lahaye 
\thanks{
The work of Jan Komenda is supported by RVO 67985840 and GA{\v C}R
grant 19-06175J.}
\thanks{Pavel \v{S}pa\v{c}ek is with Technical University Brno, Faculty of Civil Engineering, 
Veve{\v r}\'i 95, 602 00 Brno, Czech Republic, Jan Komenda is with Institute of Mathematics, Academy of Sciences of the Czech Republic, {\v Z}i{\v z}kova 22, 616 62 Brno, Czech Republic, S\'ebastien Lahaye is with LARIS, University of Angers, 62 Avenue Notre Dame du Lac, 49000 Angers,
 France.}}

\begin{document}
\maketitle \thispagestyle{empty} \pagestyle{empty}

\begin{abstract}
In this work we investigate the behavior of P-time event graphs, a class of time Petri nets with nondeterministic timing of places. Our approach is based on combined linear descriptions in both (max,+) and (min,+) semirings, where lower bounds on the state vector are (max,+)-linear and upper bounds  are (min,+)-linear.
We present necessary and sufficient  conditions for the existence of  extremal (fastest and slowest) periodic trajectories that are derived from the new description. The  results are illustrated by a realistic example of an electroplating  process.
\end{abstract}

\section{Introduction}

Timed discrete event systems arise in technological systems, where timing
of events is important in addition to logical order of events. 
P-time Petri nets are Petri nets 
with nondeterministic timing of places that have been proven useful in analysing manufacturing
systems. 
For instance, in electroplating industry \cite{Spacek99,Loiseau13}, food or chemical industry  both upper
and lower bounds on the operation times do matter. The  operation times are not given exactly, but  only by their lower and upper bounds.

P-time Petri nets with such nondeterministic
timing of places given by holding (sejourn) time intervals have been
introduced in \cite{Berthomieu}.
Timed extensions of Petri nets often admit a linear description by recurrent equations in idempotent semirings, sometimes called (max,+) and (min,+) algebras.
Such linear representations in (max,+) and (min,+) algebras are not limited to the class of timed Petri nets without  choice phenomena, called timed event graphs,  but has been extended to  1-safe  timed  Petri nets in \cite{GM99}.


We propose an approach for modeling P-time event graphs that
combines linear inequalities in both (max,+) and (min,+) algebras.
A similar idea has appeared in \cite{BH12}, but it uses the description with formal power series rather than standard state variable approach we are using here.

Event graphs with nondeterministic sojourn times of tokens in places  have been studied
using (max,+) algebra in  \cite{LH06} and \cite{Declerck}, where the emphasis has been put on their optimal control. The work in \cite{ADLM12} has also focused on control, more precisely on feedback control based on model in the (max,+)-algebra, while the upper bounds on  firing times are taken into consideration in a form of constraints.

Control of P-time event graphs can be  studied using the approach based on constraints in \cite{MAH11} and \cite{GMH17}, because some (not all due to inherent time-non determinism) trajectories of P-time event graphs can be viewed as trajectories of (max,+)-linear systems (i.e. P-timed event graphs) subject to special type of  constraints (upper bounds on dater variables).
The concept of invariant spaces for control of timed event graphs has been proposed in \cite{Katz07}
to deal with controlled invariance, which can be viewed as imposing constraint (corresponding to a subspace)  using control at all time instants, while in \cite{GMH17} the constraints are enforced only after a finite number of steps (in a steady state).

More recently, linear programming techniques based on classical linear
algebra have been used in the cycle time analysis of P-time event graphs in \cite{Loiseau13} and in \cite{Declerck14}. The approach based on classical linear
algebra has been successfully applied to the computation of extremal cycle times, namely  1-periodic trajectories, see e.g. \cite{Declerck14, Declerck18}.

In this paper we investigate  P-time event graphs
in (max,+) and (min,+) algebras.
We start by proposing new linear recurrent inequalities in both (max,+) and (min,+)-algebras describing all time constraints.
Namely, a lower bound on the state vector is expressed  by a linear inequality in the (max,+)  algebra and an upper bound on the state vector is expressed  by a linear inequality in the (min,+) algebra.
Moreover,  we obtain more precise (tighter) lower bound as well as upper bound based on
well known basic algebraic result.

This paper is a substantial extension of \cite{KS17}, where our  modeling approach combining (max,+) and (min,+)-algebras first appeared. However, (1) in \cite{KS17}  linear recurrent inequalities in both (max,+) and (min,+)-algebras have been formulated as necessary conditions and not as an equivalent description of admissible trajectories as in this paper and (2) only a straightforward application of this combined description has been presented, while in this paper we provide a detailed investigation of existence of admissible  trajectories (also known as consistency) of PTEG systems. In particular, we also present checkable necessary and sufficient conditions for existence of extremal (minimal and maximal) periodic trajectories, both 1-periodic and  p-periodic (with $p>1$).

Our algebraic approach
leads to  necessary and sufficient conditions for the existence of extremal periodic trajectories. Deriving periodical schedules for timed Petri nets has been an important topic since the 1980's, see e.g. \cite{OSW91,CC88} but we are dealing with nondeterministic timing of places. We can compute extremal $p$-periodic solutions based on  our combined (max,+) and (min,+)-algebraic recurrent inequalities.


The paper is organized as follows. Next
section  contains  algebraic preliminaries and recalls basic notions and results used throughout this paper. In Section 3  basic facts about P-time event graphs are mentioned. In section 4 we propose their description by combining linear inequalities in both (max,+) and (min,+) semirings with tighter bounds.
Necessary and and sufficient  conditions for the existence of extremal periodic trajectories are derived in section 5.
In Section 6 we illustrate our approach by an example of an electroplating industrial process, where computation of minimal and maximal extremal periodic trajectories is presented. The conclusion is proposed in Section 7.

\section{Algebraic preliminaries}\label{sec prelim}
An algebraic structure, called  idempotent semiring, also known as dioid is used in this paper. The dual (max,+) and (min,+) semirings over (extended) real numbers are recalled below together with necessary fundamental results needed in this paper.
The set of real numbers is denoted by $\mathbb{R}$ and the subset of nonnegative numbers by $\mathbb{R}_+$.

\subsection{Dioids}
We start by recalling the definition of a dioid.
\begin{definition}
An {\it idempotent semiring} (also called dioid) is a set
$\mathcal{D}$ together with two operations (addition and multiplication),
 denoted respectively $\oplus$ and $\otimes$. $\mathcal{D}$ with  
$\oplus$ forms an idempotent semigroup, i.e. $\oplus$ is commutative, associative,
has a zero element $\varepsilon$ ( $\varepsilon \oplus a=a$
for each $a \in \mathcal{D}$), and is idempotent:  $a\oplus a=a$ for each $a
\in \mathcal{D}$.  $\otimes$ is  associative, has a unit
element $e$, and distributes over $\oplus$. Moreover,
$\varepsilon$ is absorbing for $\otimes $, i.e. $\forall a\in \mathcal{D}:\;
a\otimes\varepsilon=\varepsilon \otimes a=\varepsilon$.
\end{definition}

We recall that  any idempotent semiring can be endowed 
with the natural order  defined by: $ a \preceq b
\Leftrightarrow a \oplus b=b$.  An  idempotent semiring $\mathcal{D}$ is
said to be complete
if any  subset $A\subseteq \mathcal{D}$ contains the least upper bound 
(called supremum) denoted ${\bigoplus}_{x \in A} x$, and if $\otimes$ distributes
with respect to infinite sums. In particular, in complete  idempotent
semirings there exists the greatest element of $\mathcal{D}$ given by $ T=
{\bigoplus}_{x \in \mathcal{D}} x$ . 

Typical  examples of dioids include number dioids  such
as $\mathbb{R}_{max}=(\mathbb{R}\cup \{-\infty\}, \max ,+)$ with 
addition $\oplus$ defined by  $a\oplus b=\max(a,b)$, $\varepsilon=-\infty$, and
traditional addition playing the role of multiplication, denoted
by $a\otimes b$ (or $ab$ when unambiguous). If we add $T=+\infty$
to this set, the resulting
dioid is complete and denoted by $\overline{\mathbb{R}}_{max}$.
Natural order on $\mathbb{R}_{max}$ coincides with the standard order on (extended) real numbers.
Similarly, dual dioid, denoted by 
$\mathbb{R}_{min}=(\mathbb{R}\cup \{\infty\}, \min ,+)$ with 
addition $\oplus$ defined by  $a\oplus b=\min(a,b)$, $\varepsilon=\infty$, and
traditional addition playing the role of multiplication. Natural order on $\mathbb{R}_{min}$ is the opposite of the standard order on (extended) real numbers, i.e. $c\preceq d$ iff $c\geq d$.
 If we add the supremal element $T=-\infty$
to $\mathbb{R}_{min}$, the resulting
dioid is complete and denoted by $\overline{\mathbb{R}}_{min}$. 

For $a$ in $\mathbb{R}_{max}\setminus\{-\infty\}$ or $\mathbb{R}_{min}\setminus\{+\infty\}$, 
we denote $a^{-1}$ the element such that $a\otimes a^{-1}=a^{-1}\otimes a = e (=0)$, that is $a^{-1}$ is equal to $-a$. Recall that the absorbing property of
$\varepsilon$ means that $-\infty \otimes \infty= -\infty$ in $\mathbb{R}_{max}$,
but $\infty \otimes -\infty= \infty$ in $\mathbb{R}_{min}$.


The notation $\mathbb{N}$ is reserved for the set of natural numbers with
zero.
Since both (max,+) and (min,+) semirings are used in this paper
at the same time, we fix the notation for maximum to $\oplus$
and for minimum to $\oplus'$. We emphasize that the induced
matrix multiplication in  $\mathbb{R}_{max}$ is denoted by $\otimes$ and 
 in $\mathbb{R}_{min}$ by $\otimes'$.
We have chosen this fixed notation instead of uniform context  based notation, because  we are using both semirings at the same time.

Matrix dioids are introduced in the same manner as in the
conventional linear algebra. The identity matrix of 
$\Rb_{max}^{n\times n}$ is denoted by $E$.
$E$ is defined in the same way as in the classical
algebra, i.e. it has unit elements
in the  main diagonal:  $E_{ii}=e=0$ for $i=1,\dots ,n$ and
 zero elements $\varepsilon =-\infty$ out of the main diagonal.
The identity matrix of 
$\Rb_{min}^{n\times n}$ is denoted by $E'$, where 
$$E'_{ij}=\left\{
        \begin{array}{ll}
e \;(=0), &
             \hbox{if } i=j, \\
\varepsilon \; (=\infty), &
             \hbox{ otherwise.}\\
              \end{array}
       \right. $$

The $k$-th power of a matrix $A$ will be denoted by $A^k$ if $A \in (\mathbb{R}_{max})^{n\times n}$ and by $A^{k'}$ if $A \in (\mathbb{R}_{min})^{n\times n}$. In order to avoid confusion, we will not use numbers with prime symbols such as $k'$.

In a complete dioid $\mathcal{D}$  the Kleene star operation is defined by
 $a^*=\bigoplus_{n\in \mathbb{N}} a^n,$ where by
convention $a^0=e$ for any $a\in \mathcal{D}$. 
The notation $a^+$ stands for $a^+=\bigoplus_{n\geq 1} a^n$,
i.e. $a^*=e\oplus a^+$.
For instance, in $\mathcal{D}=\Rb_{min}^{n\times n}$
we have $A^{0'}=E'$ for any matrix $A\in \Rb_{min}^{n\times n}$.
We follow the notation of \cite{Baccelli} and denote the star
operation in $\Rb_{min}^{n\times n}$ by subscript notation
$A_*=\bigoplus'_{i\in \mathbb{N}} A^{i'}$ to differentiate it from the star
of matrices over $\Rb_{max}$ (denoted  in the standard way using superscript).

\subsection{Spectral theory of matrices in $\R_{max}$}
We briefly recall the elementary facts from spectral theory of matrices in $\R_{max}$. Let us recall that
the theory for $\R_{min}$ is symmetric and readers interested in more details
and complete treatment should refer to \cite{Gau92} and \cite{BG09}.
Similarly as in classical algebra, $\lambda \in \R_{max}$ is said to be
an eigenvalue of a matrix $A\in \R_{max}^{n\times n}$ if there exists a non zero
vector (called eigenvector) $x\in \R_{max}^n$
such that
$$A\otimes x=\lambda \otimes x.$$
Spectral radius of $A\in \R_{max}^{n\times n}$ is defined similarly as
in the conventional algebra (as the maximal eigenvalue) and is denoted by
$\rho(A)$.
Spectral theory of matrices is best understood on the underlying graph, denoted $\mathcal{G}(A)$.
$A$ is called irreducible if $\mathcal{G}(A)$ is strongly connected.

We recall the concept of cyclicity of a matrix  from
\cite{Baccelli}. For any irreducible matrix $A\in \R_{max}^{n\times n}$ there exists a positive integer $d$ such that 
$A^{k+d}= \rho(A)^d \otimes A^k$ for $k$ large
enough. The minimal value of such $d\geq 1$
is called the {\it cyclicity} of $A$.
Note that this means that columns of $A^k$ are
eigenvectors of $A^d$ corresponding to  $\rho(A)^d$.
Moreover, irreducible matrices have a single eigenvalue.

Let us recall from \cite{Baccelli} that the cyclicity of a matrix over $\R_{max}$
semiring is determined by the cyclicity of the so-called critical graph of $A$.
The critical graph of $A$, denoted $\mathcal{G}^c(A)$ is a subgraph of $\mathcal{G}(A)$
consisting of nodes and arcs that belong to the so called critical circuits,
i.e. circuits of maximal weight in $\mathcal{G}(A)$.
The critical graph has in general several maximal strongly connected components
(m.s.c.c.). The cyclicity of a m.s.c.c. is the gcd (greatest common divisor)
of the lengths of all its circuits. The cyclicity $c(\mathcal{G})$ of a graph  $\mathcal{G}$ (but also of the corresponding matrix) is the lcm (least common multiple) of the cyclicities of all its m.s.c.c.
If the graph $\mathcal{G}(A)$ of a matrix itself is strongly connected as an oriented graph, i.e.  has a single maximal strongly connected component, then 
$A\in \R_{max}^{n\times n}$ is called irreducible.
Finally, we recall from \cite{Baccelli} that if both $\mathcal{G}(A)$ and $\mathcal{G}(A^d)$ are connected,
then matrix $A$ is d-cyclic.

According to the Perron-Frobenius Theorem, cf. \cite{Gau92}, the unique eigenvalue of an irreducible matrix $A\in \R_{max}^{n\times n}$
is equal to the maximum cycle mean of the graph $\mathcal{G}(A)$.
Formally, the maximum cycle mean of $A\in \R_{max}^{n\times n}$ is equal to $\bigoplus_{k=1}^n tr(A^k)^{\frac{1}{k}}$, where $tr(M)$ is the trace of matrix $M$,
i.e. the (maxplus) sum of its elements on the main diagonal.
It is to be noted that Karp algorithm is mostly used to compute
the eigenvalue (as the maximal circuit mean). 
It is obvious that if $A$ has an eigenvalue $\lambda$, then
the matrix $\lambda^{-1}\otimes A$ has the eigenvalue $e=0$.
For an irreducible matrix $A$ there exists a solution to $x=A^*\otimes x$, i.e.
$Im(A^*)\not =\emptyset$ if, and only if, $\rho(A)\leq 0$.

We recall computation of eigenvectors, which form a basis of the eigenspace 
of $A$. A set of linearly independent eigenvectors can be computed as follows.
Given a matrix $A$ with maximum circuit weight $e$ (such as $\lambda^{-1}\otimes A$), any eigenvector associated with the eigenvalue $e$ is obtained by a linear combination of $N^c(A)$
columns of the matrix $A^+$, where $N^c(A)$
denotes the number of m.s.c.c. of the critical graph $\mathcal{G}^c(A)$. More precisely,
eigenvectors are simply found as linear combinations of columns of $A^+$ with $e=0$ on the diagonal.  Note that if the i-th column of $A^+$ is denoted by $A_{.,i}^+$ then $A_{.,i}^+$ is included in a basis  for the eigenspace if $A_{i,i}^+=0$, which is equivalent to $i$ being a node
of a critical circuit of $A$.

\subsection{Fixpoint inequalities}

We need to recall the following result about solving fixpoint equations and
inequalities.
\begin{theorem} \cite{Baccelli}
\label{Baccelli}
\begin{itemize}
\item[(i)] For $a\in D$ the inequality $a\otimes x\preceq  x$ is equivalent to the
equation $x=a^*\otimes x$
\item[(ii)] Given  $a,b\in D$, the inequality
$x\succeq  a\otimes x \oplus b$
admits $a^*\otimes b$ as the least solution.
Moreover, every solution of the inequality satisfies
$x=a^*\otimes x$.
\end{itemize}
\end{theorem}
Finally,  we also need a stronger version of the above theorem, where
implication (ii) is stated as an equivalence. 
\begin{theorem} \cite{Libeaut}
\label{fixpoint}
Given $a,b\in D$, we have
$x\succeq  a\otimes x \oplus b$ if, and only if, $x\geq a^*\otimes b$ and $x=a^*\otimes x$.
\end{theorem}
We apply the above theorems to  semirings of matrices over $\R_{max}$  and $\R_{min}$. For $D=\R_{max}^{n\times n}$ it suffices to replace $\preceq$ by $\leq$ , while for $D=\R_{min}^{n\times n}$ the 
natural order is componentwise opposite order, hence $\preceq$ is replaced by $\geq$.
It follows from  (i) of Theorem \ref{Baccelli} 
that inequality $A\otimes x \leq x$ is equivalent to the
equation $x=A^*\otimes x$.
We recall the following result.
\begin{proposition}\cite{Libeaut}
Let $A^*\in \R_{max}^{n\times n}$, i.e. has finite entries. Then $\{ x \;| \; A\otimes x \leq x \}$ is generated by the columns of $A^*$. 
\end{proposition}
We  often use spaces generated by solutions of
several inequalities of type $A\otimes x \leq x$ with $x\in \R_{max}^n$, a given matrix $A\in \R_{max}^{n\times n}$, and intersection of spaces generated by several such constraints.
\begin{proposition}
\label{intersection}
 For $A,B\in \R_{max}^{n\times n}$, the system
of inequalities
\begin{eqnarray*}
A\otimes x&\leq &  x\\
B\otimes x&\leq &  x
\end{eqnarray*}
is equivalent to the single equation $x=(A\oplus B)^*\otimes x$.
\end{proposition}
{\bf Proof.} 
The system is equivalent to a single inequality $A\otimes x \oplus B\otimes x=(A\oplus B)\otimes x \leq   x$, hence the result follows from (i) of Theorem \ref{Baccelli}. 
\hfill$\square$\\
Otherwise stated, the solution space generated by 
the two fix-point inequalities above is simply the image of the star matrix
of the sum of underlying matrices that we denote as follows: $x\in Im((A\oplus B)^*)$.

\subsection{Residuation of matrix multiplication} 

We need results from residuation theory which will enable us to switch between
inequalities in $\mathbb{R}_{max}$ and $\mathbb{R}_{min}$. Since we need to work with both semirings at the same time, we do not use the concept $\preceq$ of natural order of a semiring,
but standard order on $\mathbb{R} \cup \{\pm \infty \}$. 
We recall the notation $\oplus'$  and $\otimes'$ for matrix addition and multiplication in $\mathbb{R}_{min}$ defined above.

In $\Rb_{max}$, scalar and matrix multiplications are residuated meaning that the greatest solution $X$ to the inequality $A\otimes X\leq B$ always exists. Similarly, the smallest solution $X'$ to the inequality $A'\otimes' X'\geq B'$ exists in $\Rb_{min}$. 
Recall from \cite{Gau92} that for matrices $A\in \Rb_{max}^{m\times n}$ and  $B\in \Rb_{max}^{m\times p}$
$$A\setminus B\in \Rb_{max}^{n\times p}: \; (A\setminus B)_{il}={\oplus'}_{j=1}^m A_{ji}\setminus B_{jl}.$$
Residuation over $\Rb_{max}^{n\times n}$ may then be  viewed as a dual matrix multiplication (i.e. in the (min,+)-semiring)
by the so-called conjugate matrix of $A$ in the terminology of \cite{Butkovic},
which is defined as the transpose of the matrix consisting of element-wise inverse
 of $A$, where the inversion of (max,+)-multiplication (addition) is meant. We use the residuation symbol for this matrix, i.e.
 $A^{\sharp}_{ij}=A_{ji}^{-1}$ if $A_{ji} \in \R$, $A^{\sharp}_{ij}=\infty$ if $A_{ji}=-\infty$, and  $A^{\sharp}_{ij}=-\infty$ if $A_{ji}=\infty$. We then have $A\setminus x=A^{\sharp} \otimes' x$.  Conjugate matrix  $A^{\sharp}$ is defined for any matrix on extended real numbers $\Rb \cup \pm \infty$, and 
 $A^{\sharp}$ is understood as matrix over $\Rb_{min}$ for $A$ over $\Rb_{max}$ and vice versa in this paper.

For the sake of future reference we need the following special case
of residuation from \cite{Butkovic} as a formal result.
\begin{theorem} 
\label{residuation}
For $A\in \Rb_{max}^{m\times n}$, $x\in \Rb_{max}^n$,
and $y\in \Rb_{min}^m$: 
$A\otimes x\leq y$
iff
$x\leq  A^{\sharp} \otimes' y$.
\end{theorem}
We need the properties of conjugated matrices below.
\begin{lemma} 
\label{conjugate}
\begin{itemize}
\item
For $A\in \Rb_{max}^{m\times n}$:
$\left({A^{\sharp}}\right)^{\sharp}=A$.
\item
For $A_i  \in \Rb_{max}^{m\times n}, \; i=1,\dots ,r$
$({\oplus_{i=1}^r A_i})^{\sharp}={\oplus'}_{i=1}^r A_i^{\sharp}$.
\item
For $B_i  \in \Rb_{min}^{m\times n},\; i=1,\dots ,r$
${({\oplus'}_{i=1}^r B_i)}^{\sharp}=\oplus_{i=1}^r B_i^{\sharp}$.
\item
For $A\in \Rb_{max}^{m\times n}$ and $B\in \R_{max}^{n\times p}$:
$(A\otimes B)^{\sharp}=B^{\sharp}\otimes' A^{\sharp}$.
\item
For $A\in \Rb_{max}^{n\times n}$:
${(A^*)}^{\sharp}={A^{\sharp}}_{*}$.
\item
For  $B\in \Rb_{max}^{n\times n}$
$({{(B^{\sharp})}_{*} })^{\sharp}=B^*$.
\end{itemize}
\end{lemma} 
{\bf Proof.}  
The first item is obvious from the definition of $A^{\sharp}$:
both inversion and transposition are closure mappings.
For the second claim,
$(\oplus_{i=1}^r A_i)^{\sharp}_{kl}= (\oplus_{i=1}^r A_{i})_{lk}^{-1}=
({\oplus'}_{i=1}^r A_{i}^{-1})_{lk}=({\oplus'}_{i=1}^r A_i^{\sharp})_{kl}$.
The third claim is symmetric and the proof can be obtained by interchanging
maximum and minimum operations.
The forth claim is also easy:
$(A\otimes B)^{\sharp}_{ij}=(A\otimes B)_{ji}^{-1}=
(\oplus_{k=1}^n A_{jk}\otimes B_{ki})^{-1}=
{\oplus'}_{k=1}^n A_{jk}^{-1}\otimes B_{ki}^{-1}=
({\oplus'}_{k=1}^n   A_{jk}^{-1}\otimes' B_{ki}^{-1}=
{\oplus'}_{k=1}^n B_{ik}^{\sharp}\otimes' A_{kj}^{\sharp})=
(B^{\sharp}\otimes' A^{\sharp})_{ij}$.
The second last item is as follows:
${(A^*)}^{\sharp}={\left(\oplus_{k=0}^{\infty} A^k \right)}^{\sharp}=
{\oplus'}_{k=0}^{\infty} {\left(A^k\right)}^{\sharp}={\oplus'}_{k=0}^{\infty} {(A^\sharp)}^{k'}=
{A^{\sharp}}_{*}$. 
The last item follows from the statements above.
\hfill$\square$

\section{ P-time Event Graphs} 
In this section P-time event graphs are recalled. 
 
\begin{definition} 
\label{PN}
A Petri net is a 4-tuple 
$G=(\mathcal{P},\mathcal{T},\mathcal{F},M)$, where $\mathcal{P}$ is a finite set of places, $\mathcal{T}$ is a finite set of transitions, 
$\mathcal{F}\subseteq(\mathcal{P}\times \mathcal{T})\cup (\mathcal{T}\times \mathcal{P})$ is a relation between places and transitions and $M$ is a map 
$\mathcal{P} \rightarrow \mathbb{N}$ with $M(p)$  the initial marking of place 
$p\in \mathcal{P}$. 
\end{definition}

The set $\mathcal{F}$ is composed of arcs (arrows) from places to transitions and vice versa. The notation $Pre(p)$, $Post(p)$, $Pre(a)$, and $Post(a)$ means resp. sets of upstream transitions of place $p$, downstream transitions of $p$, upstream places of transition $a$, and downstream places of transition $a$. The state (marking) of a net evolves according to usual firing rules of transitions. The firing of $a\in \mathcal{T}$ with at least one token in places from $Pre(a)$ removes one token from every place of $Pre(a)$ and adds one token to every place from $Post(a)$.





P-time Petri nets have been introduced in \cite{Khansa1996}.
We recall that event graphs are Petri nets with exactly one upstream and  one downstream transition for each place.  A P-time event graph (PTEG) is a P-time Petri net, which is an event graph.
\begin{definition} A PTEG is a 5-tuple 
$({\mathcal{P},\mathcal{T},F},M,Is)$ where $(\mathcal{P},\mathcal{T},\mathcal{F},M)$ is an event graph and \\
$Is: {\mathcal{P}}\rightarrow \mathbb{R}_+ \times \mathbb{R}_+$ is a static  interval map $p \mapsto [\tau^{min}(p),\tau^{max}(p)]$
 with $\tau^{min}(p) \leq \tau^{max}(p).$ 
\end{definition}
The firing rules of a PTEG  are as follows.
The interval $[\tau^{min}(p),\tau^{max}(p)]$ consists of a lower and an upper bound
on holding (sojourn) time for tokens in place $p$.
Every token entering place $p$ must remain there
$\tau^{min}(p)$ units of time before enabling firing of its downstream
transition. 

%

 We assume in the paper that there are no cycles without tokens in the PTEG
and the PTEG is connected.
We denote by $M^{max}$ the maximal number of initial tokens (of initial marking) in
one place.
We recall from \cite{ADL05} a simplification based on the behavioral equivalence of  P-time event graphs given by replacing every place containing $m\geq 2$ tokens by place with 1 token and the same timing and upstream
 sequence of $m-1$  transitions and places with timing $[0,0]$.  
This leads to the event graphs, where
the initial marking is bounded by 1. Hence the dimension of the state vector (i.e. the vector of dater variables used in the representation introduced in next section) for event graphs with $n$ transitions is thereby
considerably reduced from $n\times M^{max}$ to $d=n+\sum_{p\in \mathcal{P}}M(p)-p_1$, where $p_1$ is the number of places
in the PTEG  that have at most one token of the initial marking.

We complete this section with formal definitions of behavior and liveness of PTEGs. 
The semantics of PTEGs used in this paper is based on the requirement that every token entering a place $p\in \mathcal{P}$ must stay in $p$ at least $\tau^{min}(p)$ time units and at most
$\tau^{max}(p)$ time units before this token becomes available for firing of its (unique) downstream transition. However, the problem is that if this downstream transition is a synchronization transition (i.e. there exists another place say $p'\in Pre(Post(p))\subseteq \mathcal{P}$), then
the token can not be used for firing its downstream transition unless there is a token 
in $p'$ that has also already spent the required time (from the interval $[\tau^{min}(p'),\tau^{max}(p')]$). This means that the token in place $p\in \mathcal{P}$ might be forced to stay in $p$ for more time units than specified by the upper bound $\tau^{max}(p)$.
In this case we say that the token dies in $p$.
The transitions of a PTEG are fired instantaneously, which distinguishes PTEGs from T-time event graphs, where time is associated to transitions, but there are no sojourn time constraints.

It should be clear that the behavior of a PTEG is time-nondeterministic, because unlike timed event graphs, where earliest semantics yields a precise time trace (for every sequence of transition a unique duration is associated), we can only determine an interval in which the sequence of consecutive transitions can be completed.  
The trajectory of a PTEG is then  given by concrete firing times of transitions specified by a vector $x(k)$ composed of functions $x_t(k)\in \R$ for all $k=0,1,\dots$ and $t\in \mathcal{T}$,  which correspond to  dates of k-th firing of transition $t$.
 



\section{Algebraic description of P-time Event Graphs in (max,+) and (min,+) semirings }
\label{sec:modeling}
Now we present a new algebraic description of P-time event graphs that
combines both (max,+) and (min,+) descriptions.
State vector $x(k)\in \R_{max}^n$ is employed, where its component $x_i(k)$ is the date of k-th firing of transition $x_i$, also called dater function. 
 

In the proposed transformation of PTEG the places can be split into two categories according to the initial marking:
places without tokens and places with exactly one token.

From standard description of timed event graphs in $\R_{max}$
we obtain inequality of the form $x(k) \geq A\otimes x(k-1)\oplus \underline{B}\otimes x(k)$
for $k=1,2,\dots$  based on the lower bound of static intervals.
This lower bound of the sojourn time interval
$[\tau^{min}_{ji},\tau^{max}_{ji}]$ of a place with one token in the initial marking from 
transition $x_i$ (with dater variable $x_i(k)$) to 
transition $x_j$ (with dater variable $x_j(k)$) leads
to the following constraints for $k=1,2,\dots$:
\begin{eqnarray*}
x_j(k)\geq x_i(k-1)+\tau^{min}_{ji}
\end{eqnarray*}
Hence, we get $A_{ji}=\tau^{min}_{ji}$.

In a similar way,  the matrix $\underline{B}$  is obtained from the lower bound of timing 
$[\tau^{min}_{lr},\tau^{max}_{lr}]$ for the places without tokens in the initial marking from
transition $x_r$ associated with dater variable $x_r(k)$ to 
transition $x_l$ associated with dater variable $x_l(k)$. 
We have
$x_l(k)\geq x_r(k)+\tau^{min}_{lr}$ for $k=1,2,\dots$.
We see that the interval bounds $[\tau^{min}_{lr},\tau^{max}_{lr}]$   appear in $\underline{B}$ as follows:
$\underline{B}_{lr}=\tau^{min}_{lr}$.
Hence, we obtain standard (max,+)-linear inequalities
$x(k) \geq A\otimes x(k-1)\oplus \underline{B}\otimes x(k)$
for $k=1,2,\dots$ with initial condition (vector) $x(0)$.

Secondly, we derive inequalities of the form $x(k) \leq   \overline{B}\otimes' x(k) \oplus' C\otimes' x(k-1)$ \mbox{for all } k=1,2,\dots.
For the upper bound constraint associated to a place without tokens in the initial marking from
transition $x_r$  to transition $x_l$, as above we have:
\begin{eqnarray*}
x_l(k)\leq x_r(k)+\tau^{max}_{lr}
\end{eqnarray*}
Since in case $x_l$ is a  synchronization transition, we can have several constraints of this form for different upstream places without tokens and  upstream transitions $x_{r'}$, it follows that
$x_l(k)$ must be smaller than or equal to several such terms. Hence, it must be 
smaller than or equal to minimum of all such terms.
Therefore, we have constraints of the form
$x(k)\leq \overline{B}\otimes'x(k)$
with
$\overline{B}_{lr}=\tau^{max}_{lr}$.
Note that by Theorem \ref{residuation} this can be equivalently written as
$ \overline{B}^{\sharp} \otimes x(k)\leq x(k)$.
Hence, by combining this constraint with $ \underline{B}\otimes x(k)\leq x(k)$
we obtain total constraint caused by places without tokens in the equivalent form
$ [\underline{B}\oplus \overline{B}^{\sharp} ]\otimes x(k)\leq x(k)$.
We define matrix $B=\underline{B}\oplus \overline{B}^{\sharp}$ and get
$B \otimes x(k)\leq x(k)$, which improves the (max,+)-inequality above to:
$x(k) \geq A\otimes x(k-1)\oplus B\otimes x(k)$.

Finally, upper bound constraint  due to the maximal timing
$\tau^{max}_{ji}$ of a place with one token in the initial marking from 
transition $x_i$ (with dater variable $x_i(k)$) to 
transition $x_j$ (with dater variable $x_j(k)$) yields: 
\begin{equation*}
 x_j(k) \leq x_i(k-1)+\tau^{max}_{ji} \mbox{ for } k=1,2,\dots.
\end{equation*}
Again, $x_j$ is in general synchronization transition and there can be several such constraints for different $x_i$ going to $x_j$, hence we have inequality of the form\\
$x(k) \leq  C\otimes' x(k-1)$ for $k=1,2,\dots ,$.

Note that $B \otimes x(k)\leq x(k)$ can be written in (min,+) as  $x(k)\leq B ^{\sharp} \otimes' x(k) $.
Combining constraints  we get finally
(min,+) recurrent inequalities of the form
$x(k) \leq   B^{\sharp}\otimes' x(k) \oplus' C\otimes' x(k-1)$ for   $k=1,2,\dots$
with initial state vector $x(0)$.

To summarize, lower and upper bound timing constraints yield altogether
\begin{equation}
\label{ineq_maxplus}
A\otimes x(k-1)\oplus B\otimes x(k) \leq x(k) \leq   B^{\sharp}\otimes' x(k) \oplus' C\otimes' x(k-1)
\end{equation} 
for $k=1,2,\dots$,
where $A,B$ are matrices over $\R_{max}$ and  $C$ is a matrix over $\R_{min}$.
We can now formulate the concept of an admissible trajectory of a PTEG as a trajectory consistent
with its timing constraints.
\begin{definition} 
\label{admissible}
We say that a  trajectory $x(k),\; k=0,1,\dots $  is admissible for a PTEG if both lower and upper bound constraints are respected, i.e. if it satisfies the inequalities 
$$x(k) \geq A\otimes x(k-1)\oplus \underline{B}\otimes x(k) \mbox{ for } k=1,2,\dots$$ describing the lower bound constraints and the inequalities
$$x(k) \leq   B^{\sharp}\otimes' x(k) \oplus' C\otimes' x(k-1) \mbox{ for }k=1,2,\dots$$ describing the upper bound constraints.  
 \end{definition} 
We recall that timing interval
$[\tau^{min}_{ji},\tau^{max}_{ji}]$ of the place with one token in the initial marking from 
transition $x_i$  to 
transition $x_j$
appears in the matrices $A$ and $C$ as follows:
$A_{ji}=\tau^{min}_{ji}$ and $C_{ji}=\tau^{max}_{ji}$.

Timing interval  $[\tau^{min}_{lr},\tau^{max}_{lr}]$  of the place without tokens in the initial marking from 
transition $x_r$  to transition $x_j$  
appears in the matrices $\underline{B}$, $\overline{B}$ with $B=\underline{B}\oplus \overline{B}^{\sharp}$ as follows: 
$\underline{B}_{lr}=\tau^{min}_{lr}$ and $\overline{B}_{lr}=\tau^{max}_{lr}$, i.e.
$B_{lr}=\tau^{min}_{lr}$ and $B_{rl}=-\tau^{max}_{lr}$.

We emphasize that initial state $x(0)$ should be such that
$B\otimes x(0)\leq x(0)$, i.e. according to (i) of Theorem \ref{Baccelli} $B^*\otimes x(0)=x(0)$. If this condition is not satisfied, some tokens of the initial marking would violate upper bound constraints of some preceding places of the net and it would lead to dead tokens from the beginning. Note that $B^*$ must be finite (i.e. its elements are different from $+\infty$), because it does not contain positive circuits. Hence we can require that $x(0)\in Im(B^*),$
which is a very weak restriction.
We will see in the main results of the paper that stronger requirements on the initial state are needed to guarantee the existence of admissible trajectories of the PTEG.

We illustrate our approach in the following example.
\begin{example}
Consider the  P-time event graph in Figure 1.
\begin{figure}
\label{fg1}
 \centering
\scalebox{0.40}{ \includegraphics{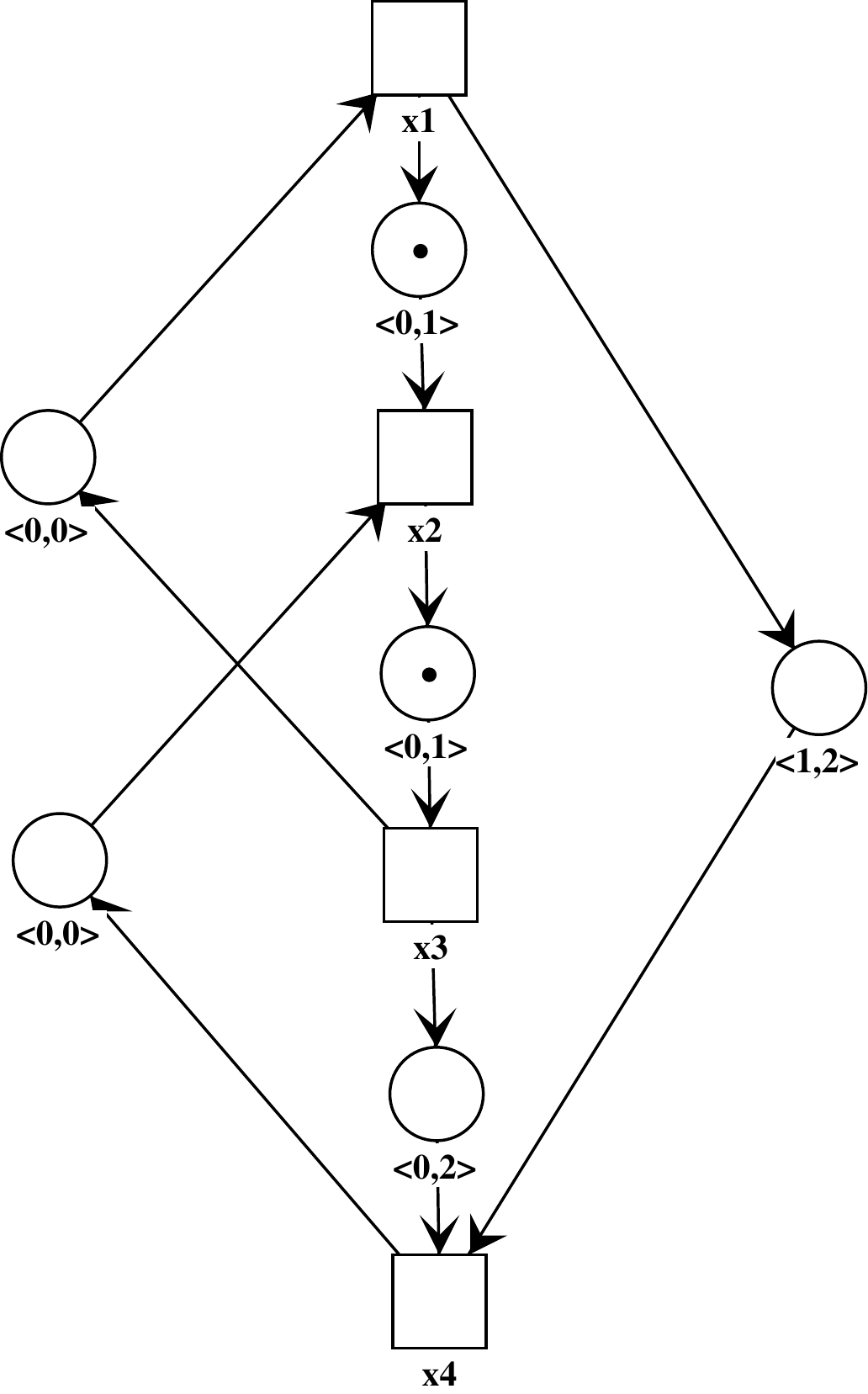}}
\caption{P-time event graph} 
\end{figure}
We obtain inequalities (\ref{ineq_maxplus}) with:
$$A=\left(
                    \begin{array}{cccc}
                      \varepsilon & \varepsilon  & \varepsilon & \varepsilon\\
                      0 &\varepsilon &  \varepsilon & \varepsilon  \\
                      \varepsilon & 0 & \varepsilon & \varepsilon \\
                      \varepsilon & \varepsilon  & \varepsilon & \varepsilon  
                    \end{array}
                  \right),
 B =\left(
                    \begin{array}{cccc}
                      \varepsilon & \varepsilon  & 0 & -2\\
                      \varepsilon & \varepsilon & \varepsilon & 0  \\
                      0 & \varepsilon & \varepsilon & -2 \\
                      1 & 0  & 0 & \varepsilon  
                    \end{array}
                  \right),$$  
$$ C=\left(
                    \begin{array}{cccc}
                      \varepsilon & \varepsilon  & \varepsilon & \varepsilon\\
                      1 &  \varepsilon & \varepsilon & \varepsilon  \\
                       \varepsilon & 1 & \varepsilon & \varepsilon \\
                      \varepsilon & \varepsilon  & \varepsilon & \varepsilon  
                    \end{array}
                  \right).
                  $$
  \end{example}

Our simultaneous description in both $\R_{max}$ and $\R_{min}$  leads to 
recurrent inequalities in the following  Proposition.

\begin{proposition}\label{combinedmodel}
A trajectory of a PTEG is admissible if and
only if it satisfies the following inequalities: 
\begin{equation} \label{final_maxplus_brief}
{\mathcal A} \otimes x(k-1) \leq x(k)\leq {\mathcal B} \otimes' x(k-1), \; k=1,2,\dots
\end{equation}
where
$$ {\mathcal A}=  B^*\otimes A\otimes B^* \mbox{ and }
{\mathcal B}=  B^{\sharp}_{*} \otimes' C \otimes' B^{\sharp}_{*}, $$
\begin{equation} \label{fix_maxplus}
\mbox{ and } x(k)=B^*\otimes x(k)  
\mbox{ for } k=0,1,\dots
\end{equation}
\end{proposition}
{\bf Proof.} 
According to the definition \ref{admissible}, an admissible trajectory is the one that satisfies inequalities (\ref{ineq_maxplus}).
In (\ref{ineq_maxplus}), inequality
$x(k) \geq  A \otimes x(k-1)\oplus B\otimes  x(k)$ for $k=1,2, \dots $, is
by Theorem \ref{fixpoint} equivalent to
$x(k) \geq  B^*\otimes A \otimes x(k-1)$ and $x(k) =  B^*\otimes  x(k)$
for $k=1,2, \dots $. Due to the remark that precedes example above
we require that the initial state $x(0)$ should be such that
$B^*\otimes x(0)=x(0)$. Hence we have $x(k-1)=B^*\otimes  x(k-1)$ for $k=1,2, \dots $.
Therefore, we get after substitution $x(k) \geq  B^*\otimes A \otimes B^*\otimes x(k-1)$ for $k=1,2,\dots $ and $x(k) =  B^*\otimes  x(k)$ for $k=0,1,\dots$.

Similarly, the other inequality in (\ref{ineq_maxplus}), 
 $x(k) \leq   C \otimes' x(k-1) \oplus' B^{\sharp}\otimes' x(k)$,  is equivalent to
$x(k) \leq  B^{\sharp}_*\otimes C^{\sharp} \otimes' x(k-1)$ and  $x(k)= B^{\sharp}_*\otimes' x(k)$
for $k=1,2, \dots $. Indeed, this follows from Theorem \ref{fixpoint} applied to (min,+) semiring (with natural order being the opposite order).
%
Note that $x(0)= B^{\sharp}_*\otimes' x(0)$ holds,
because it is equivalent to $x(0)= B^*\otimes x(0)$.

Hence, $x(k-1)= B^{\sharp}_*\otimes' x(k-1)$
for $k=1,2, \dots $ and we obtain after substitution for $k=1,2, \dots $
\begin{equation} \label{finalminplus}
x(k) \leq    B^{\sharp}_{*}\otimes'  C^{\sharp}\otimes' B^{\sharp}_{*}\otimes' x(k-1).
\end{equation}
Notice that the constraint  $x(k) = {B^{\sharp}}\otimes' x(k)$, for $k=0,1, \dots $ is not necessary (is redundant), because it is equivalent to
 $x(k)=B^*\otimes x(k)$ for  $k=0,1, \dots $.
 

Altogether, inequalities (\ref{ineq_maxplus}) can be equivalently written as follows. 
For $k=1,2, \dots $:
$$ B^*\otimes A\otimes B^*\otimes  x(k-1) \leq x(k)\leq  B^{\sharp}_{*}\otimes' C\otimes' B^{\sharp}_{*} \otimes' x(k-1),$$
$$ \mbox{ and } x(k)=B^*\otimes x(k), \; k=0,1,\dots .$$

Hence, a trajectory is admissible if
\begin{equation*}
{\mathcal A} \otimes x(k-1) \leq x(k)\leq {\mathcal B} \otimes' x(k-1), \; k=1,2,\dots
\end{equation*}
where
$$ {\mathcal A}=  B^*\otimes A\otimes B^*,
{\mathcal B}=  B^{\sharp}_{*} \otimes' C \otimes' B^{\sharp}_{*} \mbox{, and } $$
$$x(k)=B^*\otimes x(k), \; k=0,1,\dots . \hfill\square $$

We recall our assumption that there are no cycles without tokens in it. It follows that 
$B^*\in \R_{max}^{d\times d}$ has finite entries, and can be computed as the sum of the first $d-1$ powers. Without this assumption, tokens arriving in places upstream to a transition in such a cycle would die (violate the upper bound constraint).
\begin{remark}

(i) We emphasize that the obtained lower and upper bounds
proposed in Proposition \ref{combinedmodel} can be obtained in a polynomial
time (in the dimension $d$ of the square matrices $A, B$, and $C$, because the computation
of these bounds involve only matrix multiplication and the star operation (sum of the first $d-1$ powers). More precisely, the complexity of computing the bounds provided above is $O(d^3)$. 
(ii) We have not imposed any constraint on the initial state vector $x(0)$ so far.
However, it should be clear that it should verify $x_i(0)\not= -\infty$ for all $i=1,\dots ,d$, otherwise there would be dead tokens (violation of upper bound constraints already at time  $-\infty$).
Similarly, we can add a natural constraint that $x(0)\geq B\otimes x(0)$ equivalent by Theorem \ref{Baccelli} to $x(0)\in Im(B^*)$ and meaning
the (min-max) timing of places without tokens is respected.
However, we will see in the next section that one needs to restrict the initial state even further in order to guarantee the liveness, namely in addition
to $x(0)\in Im(B^*)$ we will require another condition.
\end{remark}
\begin{example}
Let us return to the running example.
According to Proposition \ref{combinedmodel},
the P-time event graph in Figure 1 
is represented by (\ref{final_maxplus_brief}) with
 $$  {\mathcal A}=B^*\otimes A\otimes B^*=
\left(
                    \begin{array}{cccc}
                      1 & 0  & 1 & 0\\
                      2 & 1 & 2 & 1  \\
                      1 & 0  & 1 & 0\\ 
                      2 & 1 & 2 & 1 
                    \end{array}
                  \right),$$

  $$  {\mathcal B}=  B^{\sharp}_{*} \otimes' C \otimes' B^{\sharp}_{*} = \left(
\begin{array}{cccc}
                      0  &-1  & 0  &-1\\
                      1 & 0 & 1 & 0  \\
                      0  &-1  & 0  &-1\\ 
                      1 & 0 & 1 & 0
                    \end{array}
                  \right).$$
\end{example}

\section{Existence of solutions}
In this section we analyze the existence of solutions to inequalities 
(\ref{final_maxplus_brief}) and provide  necessary and sufficient conditions for existence of live behaviors of PTEGs. 
\subsection{Conditions for existence of live trajectories}
It is well known that (max,+)-linear systems describing timed event graphs
(where upper bounds on holding times and firing times are missing, but
the fastest behavior called earliest functioning rule is considered)  typically reach a cyclic behavior and their cycle time
can be obtained by solving the eigenvalue problem \cite{Gau92}.
The situation is however much  more complicated for systems modeled by interval P-time event graphs, because systems described by inequalities (\ref{final_maxplus_brief}) need not have a solution at all.  
Existence of a live trajectory in a P-time event graph is equivalent to existence of a solution to system in Proposition \ref{combinedmodel}. 

First of all we note that inequalities (\ref{final_maxplus_brief}) have a solution if, and only if,
${\mathcal A} \otimes x(k) \leq {\mathcal B} \otimes' x(k)$ for $k=0,1,2\dots$.
From Theorem \ref{residuation} this is equivalent to the fix point inequality in $\R_{max}$ below:
$${\mathcal B}^{\sharp} \otimes  {\mathcal A} \otimes x(k) \leq  x(k).$$
If we add the constraint $x(k)= B^*\otimes x(k)$, then by Proposition \ref{intersection} both constraints are equivalent to 
$$({\mathcal B}^{\sharp} \otimes  {\mathcal A} \oplus B)^*\otimes x(k)= x(k).$$

The properties of matrix 
${\mathcal H}={\mathcal B}^{\sharp} \otimes  {\mathcal A}\oplus  B$ are then important for the existence of admissible trajectories of PTEG.  We see that the solution space
(admissible trajectories) is given by  image of the matrix ${\mathcal H}^*$.

First we observe that ${\mathcal H}$ is not always irreducible. However, it is obvious that if PTEG is strongly connected then
${\mathcal H}=B^* \otimes C^{\sharp} \otimes B^* \otimes  A  \otimes B^* \oplus B$ is irreducible. There is another condition that guarantees irreducibility of ${\mathcal H}$.
 \begin{proposition} 
If  there exists a path without tokens of the initial marking between any two  different transitions of the PTEG then  ${\mathcal H}$ is irreducible.
  \end{proposition}
      {\bf Proof.} 
      Let $x_i$ and $x_j$ be two different transitions of the PTEG.
Due to the meaning of matrices explained in section \ref{sec:modeling},
the graph corresponding to $B=\underline{B}\oplus  \overline{B}^{\sharp}$ is such that for every path from transition $x_u$ to transition $x_v$ in the PTEG there are paths both from $x_u$ to $x_v$ and
from $x_v$ to $x_u$ in the underlying graph of $\underline{B}\oplus  \overline{B}^{\sharp}$. This is because  matrices $\underline{B}$ and $\overline{B}$ have the same nonzero elements, and conjugate matrices have graphs with inversed arcs. Hence there exists $m\in\mathbb{N}$ such that $B^m_{ji}\not=\varepsilon$ and $B^m_{ij}\not=\varepsilon$, which means that
${\mathcal H}^m_{ij}\not= \varepsilon$ and ${\mathcal H}^m_{ji}\not= \varepsilon$
 as well. 
\hfill$\square$\\  
A necessary condition for existence of a solution  to inequalities (\ref{final_maxplus_brief}) is stated below.

\begin{proposition} \label{necond}
 Assume that $\mathcal{A}$ and $\mathcal{B}^{\sharp}$ are irreducible matrices.
If there exists a solution to inequalities (\ref{final_maxplus_brief})
then
 \begin{enumerate}
\item[(i)] $\forall n=1,2, \dots ,\max(N_{\mathcal A}, N_{\mathcal B})$: $$ [({\mathcal B}^{\sharp})^{n} \otimes  {\mathcal A}^n]\otimes x(0)\leq x(0)$$
\item[(ii)] and $\rho({\mathcal A})\leq \rho'({\mathcal B})$,
 \end{enumerate}
where $\rho'({\mathcal B})$ is the spectral radius of ${\mathcal B}$ over $\R_{min}$ (minimal mean of circuits) and $N_{\mathcal A}$, resp $N_{\mathcal B}$ are bounds due to cyclicity of ${\mathcal A}$, resp. ${\mathcal B}$.
\end{proposition}
 {\bf Proof.} 
It is obvious from monotonicity of matrix multiplication over 
both $\R_{max}$  and $\R_{min}$ that the existence of a solution to inequalities (\ref{final_maxplus_brief})
implies that $$\forall n \in \mathbb{N}, \quad A^n \otimes x(0) \leq {\mathcal B}^{n'} \otimes' x(0).$$

By repeated application of residuation ($n$ times) we have (by
Theorem \ref{residuation}):

$$\forall n \in \mathbb{N}, ({\mathcal B}^{\sharp})^{n} \otimes  {\mathcal A}^n \otimes x(0) \leq  x(0) .$$

Then by cyclicity of matrices $\mathcal{A}$ and $\mathcal{B}^{\sharp}$ there exist $N_{\mathcal A}$, $N_{\mathcal B}$, $p_{\mathcal A}$ and $p_{\mathcal B}$ such that\\
for all $n\geq N_{\mathcal{A}}$:
$\mathcal{A}^{n+p_{\mathcal A}}=\rho(\mathcal{A})^{p_{\mathcal A}} \otimes \mathcal{A}^n$ and\\
for all $n\geq N_{\mathcal{B}}$:
$({\mathcal B}^{\sharp})^{n+p_{\mathcal B}}=\rho({\mathcal B}^{\sharp})^{p_{\mathcal B}} \otimes ({\mathcal B}^{\sharp})^n$.
Note that spectral radii of $\mathcal{A}$ and ${\mathcal B}^{\sharp}$ are denoted by
$\rho(\mathcal{A})$ and $\rho({\mathcal B}^{\sharp})$.

Therefore, for $k= lcm(p_{\mathcal A},p_{\mathcal B})$ and for all $n\geq \max(N_{\mathcal A}, N_{\mathcal B})$ we obtain:
\begin{eqnarray*}
({\mathcal B}^{\sharp})^{n+k} \otimes   \mathcal{A}^{n+k}= \rho({\mathcal B}^{\sharp})^k \otimes ({\mathcal B}^{\sharp})^n \otimes
\rho(\mathcal{A})^k \otimes \mathcal{A}^n= \\
\rho({\mathcal B}^{\sharp})^k \otimes \rho(\mathcal{A})^k \otimes ({\mathcal B}^{\sharp})^n  \otimes \mathcal{A}^n.
\end{eqnarray*}
Hence, for $n\geq N:=\max(N_{\mathcal A}, N_{\mathcal B})$  we have
$$ ({\mathcal B}^{\sharp})^{n} \otimes   \mathcal{A}^n \otimes x(0)\leq x(0)  \mbox{ iff }$$ 
$$ ({\mathcal B}^{\sharp})^{N} \otimes   \mathcal{A}^N \otimes x(0)\leq x(0)  \mbox{ and }$$ 
$\rho({\mathcal B}^{\sharp})^k \otimes \rho(\mathcal{A})^k \leq 0$. This latter inequality is equivalent in classical algebra to $k(\rho({\mathcal B}^{\sharp})+ \rho(\mathcal{A})) \leq 0$. 
It clearly holds for all $k\geq 1$ if, and only if, $(\rho({\mathcal B}^{\sharp})+ \rho(\mathcal{A})) \leq 0$, i.e. $\rho({\mathcal A})\leq -\rho({\mathcal B}^{\sharp})$, which is equivalent to $\rho({\mathcal A})\leq \rho'({\mathcal B})$, because
$B$ is a matrix over $\R_{min}$, hence $\rho'({\mathcal B})=-\rho({\mathcal B}^{\sharp})$.

\hfill$\square$
  \begin{remark}
 It should be stated that condition (i) implies that  ${\mathcal A} \leq {\mathcal B}$. This follows from the case $n=1$, because all components of $x(0)$ are different from $-\infty$. This means that a simple (but too weak) necessary 
 condition for existence of a solution is ${\mathcal A} \leq {\mathcal B}$. 
 
Note that the condition of Theorem \ref{necond} can be effectively checked.
In addition, let us note that condition 
$$(i) \; [({\mathcal B}^{\sharp})^{n} \otimes  {\mathcal A}^n ]\otimes x(0)\leq x(0), 
\mbox{ for }1 \leq n\leq \max(N_{\mathcal A}, N_{\mathcal B}),$$ is by (i) of Theorem \ref{Baccelli} equivalent to 
$[({\mathcal B}^{\sharp})^{n} \otimes  {\mathcal A}^n ]^*\otimes x(0)= x(0)$.
Otherwise stated, we only need to check if $x(0)\in Im[({\mathcal B}^{\sharp})^{n} \otimes  {\mathcal A}^n]^*.$
\end{remark}





Finally we state necessary and sufficient conditions for the existence of extremal periodic trajectories. The orbits  $({\mathcal A}^n\otimes x(0)), \; n=1,2, \dots $, resp. $({\mathcal B}^{n'}\otimes' x(0)), \; n=1,2, \dots $, are natural candidates for the fastest, resp.  slowest, periodic solutions to  inequalities (\ref{final_maxplus_brief}) with a suitable choice of $x(0)$.
\begin{theorem} \label{sufcond} ~~
\begin{enumerate}
\item[(i)]
The fastest candidate solution $x(n)={\mathcal A}^n\otimes x(0)$
is a solution to inequalities (\ref{final_maxplus_brief}), i.e. it satisfies the upper bound constraints $x(n)\leq {\mathcal B}\otimes' x(n-1)$, if and only if ${\mathcal A}^{n}\otimes x(0) \in Im({\mathcal H}^*)$ for all $n =0,1,\dots$. 
\item[(ii)]
The slowest candidate solution $x(n)={\mathcal B}^{n'}\otimes' x(0)$
is a solution to inequalities (\ref{final_maxplus_brief}), i.e. it satisfies the lower bound constraints ${\mathcal A} \otimes x(n-1)\leq x(n)$,
if and only if ${\mathcal B}^{n'}\otimes' x(0) \in Im({\mathcal H}^*)$  for all $n =0,1,\dots$. 
\end{enumerate} 
\end{theorem} 
{\bf Proof.}  
Given an initial state $x(0)$, the trajectory (orbit) $x(n)={\mathcal A}^n\otimes x(0) \; n=1,2, \dots$ is a solution to  inequalities (\ref{final_maxplus_brief})
if and only if
$$x(n)={\mathcal A}^n\otimes x(0)\leq {\mathcal B}\otimes' x(n-1)={\mathcal B} \otimes' {\mathcal A}^{n-1}\otimes x(0).$$
By Theorem \ref{residuation},
$ {\mathcal A}^n\otimes x(0)\leq {\mathcal B} \otimes' {\mathcal A}^{n-1}\otimes x(0)$ is equivalent to 
${\mathcal B}^{\sharp}\otimes {\mathcal A}^n\otimes x(0)\leq {\mathcal A}^{n-1}\otimes x(0)$, i.e. ${\mathcal B}^{\sharp}\otimes {\mathcal A}\otimes {\mathcal A}^{n-1}\otimes x(0)\leq {\mathcal A}^{n-1}\otimes x(0)$, i.e.
by (i) of Theorem \ref{Baccelli}
$[{\mathcal B}^{\sharp}\otimes {\mathcal A}]^* \otimes {\mathcal A}^{n-1}\otimes x(0)={\mathcal A}^{n-1}\otimes x(0)$.
Moreover, we know from Proposition \ref{combinedmodel}
that $B\otimes x(n-1)\leq x(n-1)$, i.e. we have according to Proposition \ref{intersection} that  $x(n-1)={\mathcal A}^{n-1}\otimes x(0) \in Im([{\mathcal B}^{\sharp}\otimes {\mathcal A}\oplus B]^*)=Im({\mathcal H}^*)$ for $ n=1,2, \dots$.

The second claim can be shown similarly by observing that for $x(n)={\mathcal B}^{n'}\otimes' x(0)$ the requirement 
${\mathcal A} \otimes x(n-1)\leq x(n)$, i.e.
${\mathcal A} \otimes {\mathcal B}^{n-1}\otimes' x(0)\leq {\mathcal B}^{n'}\otimes' x(0)$ is equivalent to
$[{\mathcal B}^{\sharp}\otimes {\mathcal A}] \otimes ({\mathcal B}^{n-1}\otimes' x(0))\leq ({\mathcal B}^{n-1}\otimes' x(0))$.
\hfill$\square$\\
We will show in the next section that for a suitable choice of initial condition we can guarantee that the whole trajectory remains within the $Im({\mathcal H}^*)$
and we do not need to check the condition  for all $n$.
First, we state an important observation that explains why it is good to choose
  an eigenvector of ${\mathcal H}$ as an initial condition.
 
  \begin{proposition} If matrix ${\mathcal H}$ is irreducible then
its eigenspace  is included into 
  $Im({\mathcal H}^*)$.
    \end{proposition}
      {\bf Proof.} 
      If ${\mathcal H}$ is an irreducible matrix over $\R_{max}$,
      it has a unique eigenvalue, denoted $\lambda$.
      In any case, it is well known and recalled above how generators of
      eigenspace of ${\mathcal H}$  are constructed.
      Namely, they can be found in the columns of  $(\lambda^{-1}\otimes {\mathcal H})^+$ which have $e=0$ on the main diagonal. Since $Im({\mathcal H}^*)=Im (\lambda^{-1}\otimes {\mathcal H})^*$, ${\mathcal H}^*={\mathcal H}^+\oplus E$, and we choose as eigenvector $x$ only those columns of $(\lambda^{-1}\otimes{\mathcal H})^+$  that already have zero on the diagonal, it immediately follows that
  $x\in Im(\lambda^{-1}\otimes{\mathcal H}^+)=Im({\mathcal H}^+)$ implies that
$x\in Im({\mathcal H}^*)$.    
      \hfill$\square$\\
Let us  return again to our running example.
\begin{example}
We can check  that there is a positive circuit in ${\mathcal G}({\mathcal B}^{\sharp}\otimes {\mathcal A})$, which means that
$\rho ({\mathcal B}^{\sharp}\otimes {\mathcal A})>0$. 
Thus, $({\mathcal B}^{\sharp}\otimes {\mathcal A})^*$ is not finite and the system
of inequalities (\ref{final_maxplus_brief}) cannot have a solution.

We can see that there is a dead token in the system.
However, if the timing of the place between $x_1$ and $x_2$ 
is $[0,2]$ instead of $[0,1]$ we get the same matrix $ {\mathcal A}$, but 
$$ {\mathcal B}=  {\mathcal A}=\left(
\begin{array}{cccc}
                      1 & 0  & 1 & 0\\
                      2 & 1 & 2 & 1  \\
                      1 & 0  & 1 & 0\\ 
                      2 & 1 & 2 & 1 
                    \end{array}
                  \right).$$
                  In this case the system is time-deterministic, has a periodic behavior, and we can derive it by considering the corresponding system in 
  $\R_{max}$ or equivalently in $\R_{min}$. The matrices are irreducible, hence the
  eigenvalue is unique.
  In this case  the unique eigenvalue of matrix $ {\mathcal A}$, denoted 
$\lambda_{\mathcal A}$ coincides with unique eigenvalue $\lambda_{\mathcal B}$ of matrix $ {\mathcal B}$ computed in $\R_{min}$, i.e. $\lambda_{\mathcal A}=\lambda_{\mathcal B}=1$, because all cycle means in ${\mathcal B}=  {\mathcal A}$ are equal to $1$.
Both underlying eigenspaces have the same single generator, the shared eigenvector is
$v_{\mathcal A}=v_{\mathcal B}=(0 1 0 1)^T$. Taking $x(0)=v_{\mathcal A}$ we get the
initialization for the periodic behavior from the beginning.
\end{example}




 \subsection{Minimal and maximal periodic solutions }

Now we will investigate the existence of minimal (fastest) and maximal (slowest) periodic trajectories based on spectral theory of matrices
$\mathcal{A}$ and  $\mathcal{B}$ that generate the candidates for the extremal (minimal and maximal) behaviors.
Denote for matrices ${\mathcal A}\in \R_{max}$ and ${\mathcal B}\in \R_{min}$ 
 their (possibly non unique) eigenvalues by $\lambda_{\mathcal A}$, resp. $\lambda_{\mathcal B}$ and denote the eigenspaces corresponding to these eigenvalues by $\mathcal{P}({\mathcal A})$, resp.
$\mathcal{P}({\mathcal B})$.


Now we present formal results, which show that for extremal periodic behaviors
it is sufficient to choose the initial value from some set, hence it is sufficient to guarantee nonemptyness of this set.
We recall that  $Im(\mathcal{B}^{\sharp}\otimes \mathcal{A})^*\not=\emptyset$ if and only if
$\rho( \mathcal{B}^{\sharp}\otimes \mathcal{A})\leq 0$.
\begin{theorem} 
\label{prop1}
Consider the minimal candidate solution (i.e. the fastest periodic behavior)
 given by $x(k)=\mathcal{A}^k\otimes x(0)$ with 1-cyclic (periodic) 
matrix $\mathcal{A}$ with an eigenvalue $\lambda_{\mathcal A}$ and $\mathcal{P}({\mathcal A})$ the corresponding set of eigenvectors. Let $x(0)\in \mathcal{P}({\mathcal A})$. 
Then $x(k)$ is a solution of our system, i.e. satisfies the upper bound constraint
$x(k+1)=\mathcal{A}\otimes x(k) \leq \mathcal{B}\otimes'  x(k)$ if, and only if,
$x(0) \in  Im((\mathcal{B}^{\sharp})\otimes \mathcal{A})^*$.
\end{theorem}
{\bf Proof.}  
Let $x(0)\in {\mathcal P} ({\mathcal A})$ be such that $x(0)\in Im(\mathcal{B}^{\sharp}\otimes \mathcal{A})^*$. 
This can be written as $(\mathcal{B}^{\sharp}\otimes \mathcal{A})^*\otimes x(0)=x(0),$ which is  by (i) of Theorem \ref{Baccelli} equivalent to 
$(\mathcal{B}^{\sharp}\otimes \mathcal{A})\otimes x(0)\leq x(0),$ i.e. by Theorem \ref{residuation} this is equivalent to
$\mathcal{A}\otimes x(0)\leq \mathcal{B}\otimes' x(0)$.

Then $x(k)={\mathcal A}^k \otimes x(0)$ satisfies:
\begin{eqnarray*}
x(k)&=&{\mathcal A}^k \otimes x(0)
={\mathcal A}\otimes ({\mathcal A}^{k-1}\otimes x(0))\\
&=& {\mathcal A}\otimes (\lambda_{\mathcal A}^{k-1}\otimes x(0))
= \lambda_{\mathcal A}^{k-1}\otimes ({\mathcal A}\otimes x(0))\\
&\leq & \lambda_{\mathcal A}^{k-1}\otimes ({\mathcal B}\otimes' x(0))
= {\mathcal B}\otimes'  (\lambda_{\mathcal A}^{k-1} \otimes' x(0))\\
&=& {\mathcal B}\otimes' x(k-1).
\end{eqnarray*}
Hence, the fastest periodic behavior $x(k)={\mathcal A}^k \otimes x(0)$ satisfies
upper bound constraint.
Finally, we notice that the constraint $x(k)=B^*\otimes x(k)$  for $k>1$ is always satisfied,
because $x(k)={\mathcal A}^k \otimes x(0)$
and clearly $B^*\otimes {\mathcal A^k}=B^*\otimes (B^* \otimes A \otimes B^*)^k ={\mathcal A^k}$, because  $B^*\otimes B^*=B^*$.

Conversely, let $x(k)=\mathcal{A}^k\otimes x(0)$ is a solution of our system, i.e.
$x(k+1)=\mathcal{A}\otimes x(k) \leq \mathcal{B}\otimes'  x(k)$.
It follows from (i) in Theorem \ref{necond} that by choosing $n=1$:
$({\mathcal B}^{\sharp}) \otimes  {\mathcal A} \otimes x(0)\leq x(0)$, 
hence  by (i) of Theorem \ref{Baccelli} we have 
$x(0) \in  Im((\mathcal{B}^{\sharp})\otimes \mathcal{A})^*$.
\hfill$\square$\\
Similarly, we have the following dual claim that can be proven in a symmetric way.
\begin{theorem} 
\label{prop2}
Consider the maximal solution (i.e. the slowest periodic behavior)
given by   1-cyclic (periodic) matrix $\mathcal{B}$, i.e. $x(k)=\mathcal{B}^{k'}\otimes'x(0)$ with initialization $x(0)\in \mathcal{P}({\mathcal B})$. 
Then $x(k)$ is a solution of our system, i.e. satisfies the lower bound constraint
$\mathcal{A}\otimes x(k) \leq \mathcal{B}\otimes'  x(k)=x(k+1)$ if, and only if,
$x(0)\in Im(\mathcal{B}^{\sharp}\otimes \mathcal{A})^*$.
\end{theorem}
We emphasize that similarly as for the minimal (fastest) solution, the 
 maximal (slowest) solution automatically satisfies the constraint
 $x(k)=B^*\otimes x(k)$ for $k>1$. Indeed, it is equivalent to
 $B\otimes x(k)\leq x(k)$, which is equivalent to $ x(k)\leq B^{\sharp}\otimes' x(k)$, i.e. to $x(k)= B^{\sharp}_*\otimes' x(k)$, which is easy to see from
 $x(k)=\mathcal{B}^{k'}\otimes'x(0)$.
 Indeed, we recall that $\mathcal{B}=B^{\sharp}_{*}\otimes' C\otimes' B^{\sharp}_{*}$, hence $x(k)= B^{\sharp}_*\otimes' x(k)$ due to 
$B^{\sharp}_*\otimes' B^{\sharp}_*=B^{\sharp}_*$.
Both extremal solutions to (\ref{final_maxplus_brief}) then automatically satisfy
the additional constraint $x(k)=B^*\otimes x(k)$ for $k>1$.
We have  shown
that for 1-cyclic (periodic) extremal behaviors the state trajectory
$x(n), \; n=0,1,\dots$ remains within $Im(\mathcal{H}^*)$ whenever
$x(0)\in Im(\mathcal{H}^*)$. 

We generalize the above results to the case, where matrices $\mathcal{A}$, resp. $\mathcal{B}$ are $p$-cyclic, resp. $q$-cyclic  for some $p>1$, resp. $q>1$.
We recall that for any irreducible matrix $A\in \R_{max}^{n\times n}$ its cyclicity  is the smallest positive integer $p$ such that 
$A^{k+p}= \rho(A)^p \otimes A^k$ for $k$ large
enough. This means that columns of $A^k$ are in fact eigenvectors of
$A^p$ corresponding to its unique eigenvalue $\lambda(A^p)=\rho(A)^p$.
Let $x(k)=\mathcal{A}^k\otimes x(0)$ with $p$-cyclic (periodic) matrix $\mathcal{A}$ and initialization $x(0)\in \mathcal{P}({\mathcal A})$. 
Then we know that $\mathcal{A}^p\otimes x(0)=\lambda({\mathcal A})^p \otimes x(0)$.
We can generalize the previous results as follows.
\begin{theorem} 
\label{theorem1}
Consider the minimal candidate solution (i.e. the fastest behavior)
given by $x(k)=\mathcal{A}^k\otimes x(0)$ with $p$-cyclic (periodic) 
matrix $\mathcal{A}$ with an eigenvalue $\lambda_{\mathcal A}$ and $\mathcal{P}({\mathcal A})$ the corresponding set of eigenvectors. Let $x(0)\in \mathcal{P}({\mathcal A})$. 
Then $x(k)$ is a solution of our system, i.e. satisfies the upper bound constraint
$x(k+1)=\mathcal{A}\otimes x(k) \leq \mathcal{B}\otimes'  x(k)$ if, and only if,
$x(i) \in  Im((\mathcal{B}^{\sharp})\otimes \mathcal{A})^*$ for $i=0,1,\dots ,p-1$.
\end{theorem}
{\bf Proof.}  
Since $\mathcal{A}$ is $p$-cyclic and $x(0)\in \mathcal{P}({\mathcal A})$, we have  $\mathcal{A}^p\otimes x(0)=\lambda({\mathcal A})^p \otimes x(0)$.

We assume that
$x(i) \in Im((\mathcal{B}^{\sharp})\otimes \mathcal{A})^*$ for $i=0,1,\dots ,p-1$, which can be equivalently written as
\begin{eqnarray*}
{\mathcal A}\otimes x(0) &\leq & {\mathcal B}\otimes'  x(0), \\
{\mathcal A}^2\otimes x(0) &\leq & {\mathcal B}\otimes' [{\mathcal A}\otimes x(0)],\\
 \dots\\
{\mathcal A}^{p}\otimes x(0) &\leq & {\mathcal B}\otimes' [{\mathcal A}^{p-1}\otimes x(0)].\end{eqnarray*}
We distinguish $p$ cases depending on the modules of $k$ when divided by $p$:
 $k=pl, k=pl+1, \dots k=pl+p-1$.
Let $k=pl+r$ for some $0\leq r\leq p-1$.
Then
\begin{eqnarray*}
x(k)&=&{\mathcal A}^k\otimes x(0) ={\mathcal A}^r\otimes {\mathcal A}^{pl}  \otimes x(0)=\\
&=& {\mathcal A}^r\otimes (\lambda_{\mathcal A})^{pl} \otimes x(0)=(\lambda_{\mathcal A})^{pl} \otimes   {\mathcal A}^r \otimes x(0)=\\
&\leq &  (\lambda_{\mathcal A})^{pl} \otimes {\mathcal B}\otimes' [{\mathcal A}^{r-1}\otimes x(0)] \mbox{ assumption for i=r }\\
&=& {\mathcal B}\otimes' [{\mathcal A}^{r-1}\otimes (\lambda_{\mathcal A})^{pl} \otimes x(0)]={\mathcal B}\otimes' [{\mathcal A}^{r-1}  \otimes x(pl)]=\\
&=& {\mathcal B}\otimes'  x(pl+r-1)={\mathcal B}\otimes'  x(k-1),
\end{eqnarray*}
which was to be shown.
\hfill$\square$

The following dual result can be proven in a symmetric way.
 \begin{theorem} 
\label{prop3}
Consider the maximal candidate solution (i.e. the slowest behavior)
given by $x(k)=\mathcal{B}^{k'}\otimes' x(0)$ with q-cyclic (periodic) 
matrix $\mathcal{B}$ that has an eigenvalue $\lambda_{\mathcal B}$ and $\mathcal{P}({\mathcal B})$ the corresponding set of eigenvectors. Let $x(0)\in \mathcal{P}({\mathcal B})$. 
Then $x(k)$ is a solution of our system, i.e. satisfies the lower bound constraint
$x(k+1)=\mathcal{B}\otimes' x(k) \geq \mathcal{A}\otimes  x(k)$ if, and only if,
$x(i) \in  Im((\mathcal{B}^{\sharp})\otimes \mathcal{A})^*$ for $i=0,1,\dots ,q-1$.
\end{theorem}

\section{Example of an electroplating line}
Below an industrial example is presented, where our combined (max,+)/(min,+) modeling approach with improved lower and upper
bounds is applied. This example of an electroplating line is taken from
\cite{Loiseau13}. The authors have proposed therein a linear algebraic approach
(in classical linear algebra) based on linear programming.
The P-time event graph model of the electroplating line is depicted on Figure 2. We emphasize that
$x_9(k)$ of the state vector is actually $x_{41}(k)$ of the PTEG, which stems
from the extension of the state vector according to the simplification procedure that is described above.

\begin{figure}
\label{fg2}
 \centering
\scalebox{0.4}{ \includegraphics{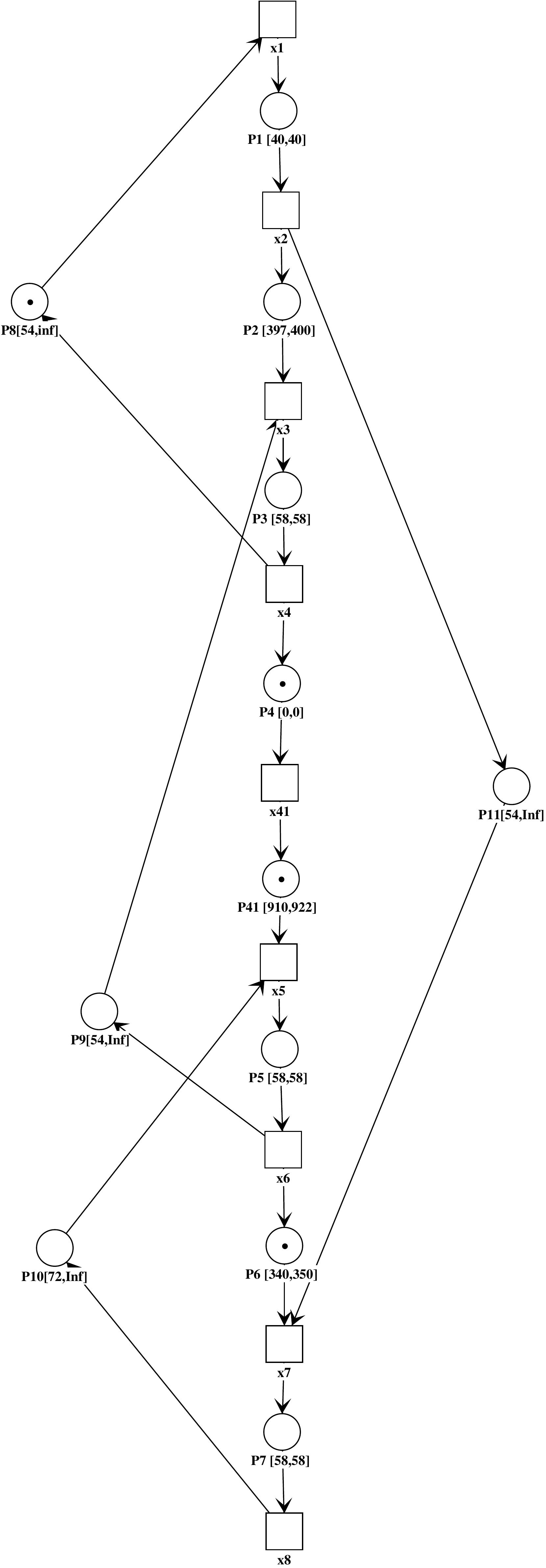}}
\caption{Model of an electroplating line} 
\end{figure}

According to Proposition \ref{combinedmodel}
we obtain the following first order inequalities on the state vector:
 $${\mathcal A} \otimes x(k-1) \leq x(k)\leq {\mathcal B} \otimes' x(k-1)$$
with     
$${\mathcal A}=  \left(
\begin{array}{ccccccccc}
549  & 509   & 112 & 54 & 224
& 166 & 354   & 296 & 582\\
 589  & 549   & 152 & 94 & 264
& 206  & 394   & 336 & 622\\
986 & 946   & 549 & 491 & 661
& 603  & 791  & 733 & 1022\\
1044  & 1004   & 607 & 549 & 719
& 661  & 849   & 791 & 1080\\
773  & 773  & 336 & 278 & 528
& 470  & 658  & 600 & 910\\
831  & 791  & 394 & 336 & 586
& 528  & 716   & 658 & 968\\
643  & 603   & 206 & 148 & 398
& 340  & 528   & 470 & 676\\
701  & 661   & 264 & 206 & 456
& 398 & 586  & 528 & 734\\
495  & 455   & 58 &0  & 170
&112  & 300   & 242 & -\infty
                    \end{array}
                  \right)$$ 
                  and 
$${\mathcal B}=  \left(
\begin{array}{ccccccccc}
642  & 602   & 202 & 144 & 314
& 256  & 548   & 490 & 698\\
682  & 642  & 242 & 184 & 354
& 296  & 588  & 530 & 738\\
1082  & 1042   & 642 & 584 & 754
& 696  & 988   & 930 & 1138\\
1140 &  1100   & 700 & 642 & 812
& 754 & 1046   & 988 & 1196\\
970  & 930   & 530 & 472 & 642
& 584  & 876   & 818 & 922\\
1028  & 988   & 588 & 530 & 700
& 642  & 934   & 876 & 980\\
736  & 696   & 296 & 238 & 408
& 350  & 642   & 584 & 792\\
794  & 754   & 354 & 296 & 466
& 408 & 700   & 642 & 850\\
498 & 458   & 58 & 0  & 274
& 216  & 404   & 346 & \infty
                    \end{array}
                  \right).$$                  
                  

It can be shown that
 ${\mathcal A}=B^*\otimes A\otimes B^*$ is a 
1-cyclic matrix over $\R_{max}$ and that ${\mathcal B}=B^{\sharp}_{*}\otimes'  
C \otimes' B^{\sharp}_{*}$  is a  3-cyclic matrix over $\R_{min}$.
Moreover, ${\mathcal A}$ is irreducible, i.e. it has a unique eigenvalue equal to
$\lambda_{\mathcal A}=549$.
Eigenvectors corresponding to $\lambda_{\mathcal A}$ can be computed in a standard way using the matrix 
$(-\lambda_{\mathcal A}\otimes{\mathcal A})^+$.
Namely, we are looking for the columns of the above matrix having $e=0$ on the diagonal.
There is a single independent eigenvector, e.g. the first column
$x_{\mathcal A}=(0\; 40 \;437 \;495\; 307\; 365\; 156\; 214\; -54)^T$. 
We can check that the 1-periodic fastest behavior corresponding 
to $x(0)=x_{\mathcal A}$ and driven by the matrix ${\mathcal A}$ (meaning $x(k)={\mathcal A}\otimes x(k-1)$)
satisfy the upper bound. Indeed, according to Theorem \ref{prop1}
it is sufficient to verify that 
\begin{equation*}
{\mathcal A}\otimes x_{\mathcal A}\leq {\mathcal B}\otimes' x_{\mathcal A}.
\end{equation*}  
We emphasize that the eigenvector with negative entries is not suitable for
practical use, but we rather choose a suitable multiple of $x_{\mathcal A}$,
which in $(max,+)$ setting simply means that a given (here positive) number $m\in \mathbb{R}$ can be added to all its components and $m\otimes  x_{\mathcal A}$
is also an eigenvector. In this example we can take
$54\otimes x_{\mathcal A}$ instead of $x_{\mathcal A}$.

Similarly, we can compute the minimal cycle mean
(generalized eigenvalue) corresponding to the upper bound matrix ${\mathcal B}$
that is 3-cyclic.
It is equal to $\lambda_{\mathcal B}=578$ and from the matrix 
$(-\lambda_{\mathcal B}\otimes'{\mathcal B})_+$ we compute a single independent vector
corresponding to 3-cyclic behavior, e.g.
$$x_{\mathcal B}=(0\; 40 \;440 \;498\; 264\; 322\; 94\; 152\; -80)^T.$$
Let us now consider the 3-cyclic slowest behavior corresponding
to $x(0)=x_{\mathcal B}$ and driven by the matrix ${\mathcal B}$  (i.e. $x(k)={\mathcal B}\otimes' x(k-1)$).
In order to ensure that underlying 3-cyclic maximal behavior initialized
by taking $x(0)=80\otimes x_{\mathcal B}$ (to have all components non negative)
satisfy the lower bound constraint, according to Theorem \ref{prop3} we only have to check if $x(i) \in  Im((\mathcal{B}^{\sharp})\otimes \mathcal{A})^*$ for $i=1,2,3$. Since $x(i)=\mathcal{B}^{i'}\otimes' x(0)$, it is equivalent to check inequalities (\ref{condition1})-(\ref{condition3}):
\begin{eqnarray}
\label{condition1}
{\mathcal A}\otimes x_{\mathcal B} &\leq & {\mathcal B}\otimes' x_{\mathcal B}, \\
{\mathcal A}\otimes [B\otimes' x_{\mathcal B}] &\leq & {\mathcal B}^{2'}\otimes' x_{\mathcal B}, \\
\label{condition3}
{\mathcal A} \otimes [B^{2'}\otimes' x_{\mathcal B}] &\leq & {\mathcal B}^{3'}\otimes' x_{\mathcal B}.\end{eqnarray}
In this example all the three inequalities are satisfied.
Again, due to 3-cyclicity of matrix ${\mathcal B}$ we have from Theorem \ref{prop3} that 
the  trajectory 
$x(k)={\mathcal B}^{k}\otimes x_{\mathcal B}$  is a solution, i.e.
$${\mathcal A}\otimes x(k-1) \leq x(k)$$ for all $k>3$ as well.
 
Finally we point out that the two extreme cycle means coincide with those obtained using linear programming in \cite{Loiseau13}. There is a small difference between the computed initializations that is due to the fact that
our approach is purely algebraic, while using linear programming some rounding errors can occur. We have preferred algebraic
 computation in $\Rb_{max}$ and $\Rb_{min}$ semirings with additional benefits such as
 polynomial time complexity (not always guaranteed when using linear programming).
The negative numbers occur in the state component that is fictive, i.e. stemming from extension of state vector due to two tokens of initial marking in one of the places. In practice there are two possibilities how to avoid the negative components: either we multiply all components by (add to them in the conventional algebra)  the inverse of this negative number (while remaining in the same space) or we can also ignore the fictive component.

Another difference between our paper and \cite{Loiseau13} is that the authors
investigate the problem of reaching the asymptotic (cyclic) regime by a controller that is computed by solving a linear program, while we can simply 
choose the initial vector in the eigenspace of a matrix and guarantee (cf. Theorems 20 and 21) that the cyclic regime is reached from the very beginning. 
 
%
\vskip 5mm
    
\section{Conclusion}
In this paper a new algebraic description of  behaviors of P-time 
event graphs has been derived. It is based on (max,+)-linear recurrent inequalities 
describing the lower bound on the state vector and  
on (min,+)-linear recurrent inequalities 
describing the upper bound on the state vector.
The proposed inequalities are tighter than the known ones, e.g.
the lower bound is improved from $B^*\otimes A$  to $B^*\otimes A\otimes B^*$
and similarly for the upper bound, which leads to new results about existence of cyclic behaviors.

We  have obtained necessary and sufficient conditions
for the existence of $p$-cyclic extremal trajectories that guarantee that the problem of dead tokens does not occur.


Our approach is illustrated by a realistic industrial example based on a model of  an  electroplating process taken from \cite{Loiseau13}. We advocate purely algebraic techniques for computing fastest and slowest cyclic behaviors rather than linear programming techniques used in the reference.

We believe that our new approach will also be useful
in other performance evaluation and control problems for PTEGs.
In particular, we plan to apply our algebraic description of solution space to
problems of switching between different (cyclic) regimes given by linearly independent eigenvectors, which is useful in reconfiguration of manufacturing systems modeled by PTEGs.

\end{document}